\RequirePackage{fix-cm}
\documentclass[onecolumn,epjc3]{svjour3}  
\smartqed  
\RequirePackage{graphicx}
\usepackage{amsmath}
\usepackage{amssymb}
\usepackage{centernot}
\usepackage{dsfont}
\usepackage{color}
\newcommand{\nn}{\nonumber}
\newcommand{\uvec}{\boldsymbol}
\newcommand{\ud}{\mathrm{d}}
\newcommand{\fsl}[1]{{\centernot{#1}}}
\journalname{Eur. Phys. J. C}
\begin{document}

\title{Potential linear and angular momentum in the scalar diquark model}

\author{David Arturo Amor-Quiroz\thanksref{e1,addr1}
        \and
        Matthias Burkardt\thanksref{e2,addr2} 
        \and
        William Focillon\thanksref{e3,addr1} 
        \and
        C\'edric Lorc\'e\thanksref{e4,addr1} 
}

\thankstext{e1}{e-mail: arturo.amor-quiroz@polytechnique.edu}
\thankstext{e2}{e-mail: burkardt@nmsu.edu}
\thankstext{e3}{e-mail: william.focillon@polytechnique.edu}
\thankstext{e4}{e-mail: cedric.lorce@polytechnique.edu}


\institute{CPHT, CNRS, Ecole Polytechnique, Institut Polytechnique de Paris, Route de Saclay, 91128 Palaiseau, France \label{addr1}           \and 
 Department of Physics, New Mexico State, University, Las Cruces, New Mexico 88003, USA \label{addr2}
}

\date{Received: date / Accepted: date}

\maketitle

\begin{abstract}
We present an analytic two-loop calculation within the scalar diquark model of the potential linear and angular momenta, defined as the difference between the Jaffe-Manohar and Ji notions of linear and angular momenta. As expected by parity and time-reversal symmetries, a direct calculation confirms that the potential transverse momentum coincides with the Jaffe-Manohar (or canonical) definition of average quark transverse momentum, also known as the quark Sivers shift. We examine whether initial/final-state interactions at the origin of the Sivers asymmetry can also generate a potential angular momentum in the scalar diquark model.
\keywords{Nucleon structure \and Angular momentum \and Scalar diquark model}
\end{abstract}

\section{Introduction}
\label{Intro}

One of the main challenges of hadronic physics is to fully understand the structure and characteristics of nucleons through the measurement of their structure functions. Of particular interest is the origin of the nucleon spin, a key question at the core of the experimental program of the future Electron-Ion Collider (EIC) at the Brookhaven National Laboratory~\cite{Accardi:2012qut,Proceedings:2020eah}. This requires a proper decomposition of the nucleon total angular momentum (AM) into orbital motion and intrinsic spin of its constituents. 
The most common decompositions of AM are the Jaffe-Manohar (JM)~\cite{Jaffe:1989jz} and Ji~\cite{Ji:1996ek} decompositions, which are based on two different notions of orbital angular momentum (OAM). Variations and extensions of these have been proposed and discussed at length in the literature. For further details, we refer the interested reader to the reviews by Leader and Lorc\'e~\cite{Leader:2013jra}, and Wakamatsu~\cite{Wakamatsu:2014zza}.

For the purpose of the present work, it suffices to recall that the Ji decomposition is based on the kinetic momentum defined by the covariant derivative $D_\mu=\partial_\mu-ig A_\mu$, whereas the JM decomposition is based on the canonical momentum $\partial_\mu$ defined in the light-front (LF) gauge $A^+=(A^0+A^3)/\sqrt{2}=0$. More specifically, we are interested in the \emph{difference} between JM and Ji quark OAM, known as potential AM~\cite{Wakamatsu:2010qj}
\begin{equation}
L^z_\text{pot}\equiv L^{q,z}_\text{JM}-L^{q,z}_\text{Ji}.    
\end{equation}
From a physical point of view, this potential AM has been interpreted as the accumulated change in OAM experienced by the struck quark due to the color Lorentz forces as it leaves the target in high-energy scattering processes~\cite{Burkardt:2012sd}. 

Similarly, the potential transverse\footnote{Since $A^+=0$ in the JM decomposition, there is no difference in the longitudinal LF momentum $k^+_\text{pot}=k^{q,+}_\text{JM}-k^{q,+}_\text{Ji}=0$.} momentum (TM) is defined as the difference between JM and Ji notions of the quark transverse momentum inside the nucleon~\cite{Wakamatsu:2010qj}
\begin{equation}
\uvec k_{\perp,\text{pot}}\equiv \uvec k^q_{\perp ,\text{JM}}-\uvec k^q_{\perp,\text{Ji}}.    
\end{equation}
This potential TM can be related to the Sivers shift (as justified with more detail in Section~\ref{Discussion}), which is the non-vanishing average parton transverse momentum in a transversely polarized target resulting from the Sivers mechanism~\cite{Burkardt:2003yg,Burkardt:2004ur}.

On the one hand, both the JM and Ji OAM have been computed in QED at one-loop order and found to be same~\cite{Ji:2015sio}, leading therefore to the conclusion that potential OAM vanishes at that order. On the other hand, a significant non-vanishing potential AM has been reported in recent lattice calculations~\cite{Engelhardt:2017miy,Engelhardt:2020qtg}, a numerical study of the renormalization scale dependence of $L^z_\text{pot}$ at one-loop order has been shown to be non-trivial owing to the mixing with a tower of operators provided that the corresponding initial values do not vanish~\cite{Hatta:2019csj}, and the potential AM has appeared to play an essential role in the context of the Landau problem~\cite{Wakamatsu:2017isl}. All these hint at a non-vanishing potential AM.\\

The aim of the present work is to investigate within an explicit model calculation whether the difference between Ji and JM definitions of OAM does appear at two-loop level.

Both the potential TM and potential AM can in principle be experimentally observed. A comparison of their magnitudes is motivated by Burkardt's proposal of a lensing mechanism due to soft gluon rescattering in deep-inelastic and other high-energy scattering~\cite{Burkardt:2002ks,Burkardt:2003uw}. Originally, such mechanism describes the asymmetry in transverse momentum space (i.e. Sivers effect) as originating from an asymmetry in impact-parameter space (due to the orbital motion of partons) convoluted with a lensing function that accounts for the effect of attractive initial/final state interactions (ISI/FSI). Even if the notion of a lensing function is intuitive and is supported by some model calculations~\cite{Burkardt:2003uw,Burkardt:2003je,Bacchetta:2011gx,Gamberg:2009uk}, the ability of factorizing it depends on strong assumptions that cannot be strictly satisfied in QCD~\cite{Pasquini:2019evu}. Nevertheless, a non-zero Sivers function necessarily requires a non-vanishing orbital motion of the partons. The existence of a non-vanishing potential TM hints towards a non-vanishing potential AM, since both quantities find their common origin in the ISI/FSI. It is therefore interesting to compare these quantitatively within the same approach.

In the present paper, we compute both the potential TM and potential AM in the framework of a simple scalar diquark model (SDM) of the nucleon at order $\mathcal{O}(\lambda^2 e_q e_s)$ in perturbation theory. The model is based on the assumption that the nucleon splits into a quark and a scalar diquark structure, which are regarded as elementary fields of the theory~\cite{Brodsky:2002cx}. More details on the model are presented in~\ref{AppSDM}.

Despite of being a simple model, the SDM provides analytic results that have broadly been explored in the literature. Additionally, it has the feature of maintaining explicit Lorentz covariance. For these reasons, we believe that the SDM presents a good framework for providing an estimate of the magnitude of potential AM with respect to the magnitude of potential TM. The effect of introducing a vector diquark on said observables is beyond the interest of the present document. \\

The manuscript is organized as follows. We start with a reminder of the gauge-invariant definition of canonical momentum. In Section~\ref{Discussion} we define and compute the potential TM of an unpolarized quark in a transversely polarized nucleon.
Thereafter we report the potential AM in Section~\ref{PotOAM} for a quark in a longitudinally polarized nucleon. Finally, we summarize our results in Sec.~\ref{Conclusions}. Some details about the two-loop calculations are collected in Appendices.

\section{Gauge-invariant definition of canonical momentum}
\label{GIdef}

The canonical momentum refers to the partial derivative $\partial_\mu$. Since this is not a gauge covariant operator, Jaffe and Manohar considered it only in a particular gauge, namely the LF gauge $A^+$ well-suited for the description of high-energy scatterings. Since the JM decomposition is defined in a fixed gauge, doubts were expressed concerning the measurability of its individual terms, in particular of the JM quark and gluon OAM.

A decade ago, a paper by Chen \emph{et al.}~\cite{Chen:2008ag} shook the nucleon spin community by showing that the canonical decomposition of angular momentum can be written in a gauge-invariant way. To achieve this goal, one has to separate the gauge potential into so-called \emph{pure-gauge} and \emph{physical} (or better dynamical) parts~\cite{Wakamatsu:2010qj,Chen:2008ag,Wakamatsu:2010cb} 
\begin{equation}\label{Chenetaldec}
A_\mu = A^\text{pure}_\mu+A^\text{phys}_\mu.
\end{equation}
By definition, $A^\text{pure}_\mu$ has a vanishing field strength
\begin{equation}
F^\text{pure}_{\mu\nu}\equiv\partial_\mu A^\text{pure}_{\nu}-\partial_\nu A^\text{pure}_{\mu}-ig\,[A^\text{pure}_{\mu},A^\text{pure}_{\nu}]= 0, 
\end{equation}
and behaves as a connection
\begin{equation}
A^\text{pure}_\mu(x)\mapsto \widetilde A^\text{pure}_\mu(x)=U(x)\left[A^\text{pure}_\mu(x)+\frac{i}{g}\,\partial_\mu\right]U^{-1}(x)
\end{equation}
under gauge transformations. It can then be used to define a new covariant derivative
\begin{equation}
D^\text{pure}_\mu=\partial_\mu-ig A^\text{pure}_\mu.
\end{equation}
It is nothing but the gauge-invariant version of the canonical momentum operator since it satisfies the standard canonical commutation relation
\begin{equation}
[D^\text{pure}_\mu,D^\text{pure}_\nu]=-igF^\text{pure}_{\mu\nu}=0,
\end{equation}
and it generates the gauge-covariant translations of the fields (in the fundamental representation). The dynamical degrees of freedom of the gauge potential are encoded in $A^\text{phys}_\mu$ which behaves covariantly
\begin{equation}
A^\text{phys}_\mu(x)\mapsto \widetilde A^\text{phys}_\mu(x)=U(x)A^\text{phys}_\mu(x)U^{-1}(x)
\end{equation}
under gauge transformations. These are responsible for the non-vanishing of the field strength
\begin{equation}
F_{\mu\nu}\equiv \mathcal D^\text{pure}_\mu A^\text{phys}_{\nu}-\mathcal D^\text{pure}_\nu A^\text{phys}_{\mu}-ig\,[A^\text{phys}_{\mu},A^\text{phys}_{\nu}],
\end{equation}
where $\mathcal D^\text{pure}_\mu A^\text{phys}_\nu=\partial_\mu A^\text{phys}_\nu-ig\,[A^\text{pure}_\mu,A^\text{phys}_\nu]$ is the pure-gauge covariant derivative in the adjoint representation. Note that since $A^\text{pure}_\mu$ is a pure-gauge field, there exists a gauge where $A^\text{pure}_\mu=0$ and hence $A_\mu=A^\text{phys}_\mu$. We will refer to this gauge as the \emph{natural} gauge.

The separation of the gauge potential~\eqref{Chenetaldec} is however not unique~\cite{Lorce:2012rr,Lorce:2013gxa,Lorce:2013bja} and requires an additional condition to define unambiguously $A^\text{phys}_\mu$. Such condition is akin to a standard gauge condition except that it is imposed only on $A^\text{phys}_\mu$ and not on the full $A_\mu$. This amounts to defining $A^\text{phys}_\mu$ as a certain non-local functional of $A_\mu$. Owing to the gauge principle, no gauge is more fundamental than the others. Accordingly, no separation of the gauge potential~\eqref{Chenetaldec} is more fundamental than the others\footnote{Otherwise the natural gauge would appear to be more fundamental than the other gauges.}. However, when dealing with a particular problem some separation may appear more \emph{convenient} to use. This is the case of high-energy scatterings where a particular spatial direction (identified with the $z$-direction) is determined by the kinematics. In this context, the physical part reads~\cite{Boer:1997bw,Boer:2003cm,Hatta:2011ku,Lorce:2012ce}
\begin{equation}\label{AphysLF}
A^\mu_{\text{phys},\pm}(x)=\int^{x^-}_{\pm\infty^-}\ud y^-\,\mathcal W_\text{LF}(x^-,y^-;\tilde{\uvec x})\,F^{+\mu}(y^-,\tilde{\uvec x})\,\mathcal W_\text{LF}(y^-,x^-;\tilde{\uvec x}),
\end{equation}
where the LF Wilson line is defined by the following path-ordered exponential
\begin{equation}
\mathcal W_\text{LF}(x^-,y^-;\tilde{\uvec x})\equiv\mathcal P\left[\exp\left(ig\int_{y^-}^{x^-}\ud z^- A^+(z^-,\tilde{\uvec x})\right)\right].
\end{equation}
For convenience, we introduced the notation $\tilde{\uvec x}=(x^+,\uvec x_\perp)$ and the LF components $x^\pm=(x^0\pm x^3)/\sqrt{2}$. The condition $A^+_\text{phys}=0$ is not sufficient for defining unambiguously the physical contribution, since it does not fix $x^-$-independent contributions. This residual freedom is fixed by adding a boundary condition. The advanced (retarded) boundary condition corresponds to setting $+\infty^-$ ($-\infty^-$) in Eq.~\eqref{AphysLF}. More generally, we can consider any intermediate boundary condition
\begin{equation}
A^\mu_{\text{phys},\eta}(x)=\tfrac{1+\eta}{2}\, A^\mu_{\text{phys},+}(x)+\tfrac{1-\eta}{2}\,A^\mu_{\text{phys},-}(x)
\end{equation}
with\footnote{The restriction over $\eta$ ensures that the path remains within the slab of width $\ud x$.} $\eta\in[-1,1]$. The particular value $\eta=0$ is quite popular and corresponds to the antisymmetric boundary condition.

Interestingly, one can see the expression in Eq.~\eqref{AphysLF} as an infinitesimally thin and (semi)-infinitely long Wilson loop. Indeed, the parallel transport from a point $x$ to an infinitesimally closed point $x+\ud x$ is defined by the Wilson line $\mathcal W_\mathcal{C}(x+\ud x,x)=\mathds 1+igA^\mathcal{C}_\mu(x)\ud x^\mu$, where $\mathcal C$ is some path connecting the two points. As long as the path remains local, i.e. in the infinitesimal neighborhood of $x$, the connection does not depend on $\mathcal C$ and we simply write $A_\mu(x)$. Typically one associates $A_\mu(x)$ with a straight line between $x$ and $x+\ud x$. In this geometric picture, the field strength describes a closed infinitesimal square $\mathcal W_\Box(x)=\mathds 1+igF_{\mu\nu}(x)\ud x^\mu \ud y^\nu$. When the path between $x$ and $x+\ud x$ is non-local, the connection retains its path dependence. We can see $A^\mathcal{C}_\mu$ as an explicit realization of $A^\text{pure}_\mu$. By definition $A^\text{phys}_\mu=A_\mu-A^\text{pure}_\mu$ corresponds therefore to a closed non-local path. In the context of high-energy scatterings, the natural paths to consider are non-local in the LF direction $x^-$. We can then understand Eq.~\eqref{AphysLF} as the breaking up of an infinitesimally thin Wilson loop extending from $x^-$ to $\pm\infty^-$ into an infinite sum of parallel-transported infinitesimal squares. Changing boundary conditions corresponds to adding a gauge-invariant contribution proportional to
\begin{align}
&A^\mu_{\text{phys},+}(x)-A^\mu_{\text{phys},-}(x)=\nn\\
&\qquad-\int^{+\infty^-}_{-\infty^-}\ud y^-\,\mathcal W_\text{LF}(x^-,y^-;\tilde{\uvec x})\,F^{+\mu}(y^-,\tilde{\uvec x})\,\mathcal W_\text{LF}(y^-,x^-;\tilde{\uvec x}),    
\end{align}
i.e. an infinitesimally thin Wilson loop extending from $-\infty^-$ to $+\infty^-$. In general $A^\mu_{\text{pure},\eta}(x)$ is associated with a flaky (or s-like\footnote{The s and \reflectbox{s} shapes are equivalent.}) line which reduces to a staple when $\eta=\pm 1$.

\section{Average transverse momentum of unpolarized quarks inside a transversely polarized nucleon}
\label{Discussion}

We are interested in the average TM of quarks inside a nucleon. For the kinetic momentum, it is given by the following expectation value\footnote{For convenience, we omit the label $q$ indicating that we are considering the quark contribution only.} at equal LF time $r^+=0$
\begin{align}
\langle k^i_\perp\rangle_\text{Ji}&=\frac{\langle P,S|\int\ud^4 r\,\delta(r^+)\,\overline\psi(r)\gamma^+\tfrac{i}{2}\overset{\leftrightarrow}{D}\!\!\!\!\!\phantom{D}^i_\perp\psi(r)|P,S\rangle}{\langle P,S|P,S\rangle}\nn\\
&=\frac{i}{4P^+}\,\langle P,S|\overline\psi(0)\gamma^+\overset{\leftrightarrow}{D}\!\!\!\!\!\phantom{D}^i_\perp\psi(0)|P,S\rangle,
\end{align}
where $i=1,2$ is a transverse index, $\overset{\leftrightarrow}{D}\!\!\!\!\!\phantom{D}_\mu=\overset{\rightarrow}{D}\!\!\!\!\!\phantom{D}_\mu-\overset{\leftarrow}{D}\!\!\!\!\!\phantom{D}_\mu$ is the symmetric covariant derivative, and a translation of the fields to the origin has been used in the second line. The nucleon with mass $M$ has four-momentum $P^\mu$ and covariant LF polarization $S^\mu$ satisfying $P\cdot S=0$ and $S^2=-1$. Its explicit form is given by
\begin{equation}
S^\mu=[s_z \frac{P^+}{M},-s_z \frac{P^-}{M}+\frac{\uvec P_\perp\cdot\uvec S_\perp}{P^+},\uvec S_\perp]
\end{equation}
with $\uvec S_\perp=\vec s_\perp+\frac{\uvec P_\perp}{M}\,s_z$ and $\uvec s$ the usual rest-frame spin vector. For convenience, we work in the standard symmetric LF frame defined by $\uvec P_\perp=\uvec 0_\perp$. 
For the canonical momentum, we have in a similar way
\begin{equation}
\langle k^i_\perp\rangle_{\text{JM},\eta}=\frac{i}{4P^+}\,\langle P,S|\overline\psi(0)\gamma^+\overset{\leftrightarrow}{D}\!\!\!\!\!\phantom{D}^i_{\text{pure},\eta,\perp}\psi(0)|P,S\rangle.
\end{equation}
We left here the explicit dependence on the choice of the boundary condition. The potential TM is then given by
\begin{equation}
\label{defkpot}
\langle k^i_\perp\rangle_{\text{pot},\eta}=\frac{-g}{2P^+}\,\langle P,S|\overline\psi(0)\gamma^+A^i_{\text{phys},\eta,\perp}(0)\psi(0)|P,S\rangle.
\end{equation}

The theory being invariant under discrete space-time symmetries, the combined parity and time-reversal transformation (\textsf{PT}) implies that\footnote{One can use for example the general parametrization for the matrix elements of the energy-momentum tensor obtained in~\cite{Lorce:2015lna}.} 
\begin{align}
\langle k^i_\perp\rangle_\text{Ji}&=-\langle k^i_\perp\rangle_\text{Ji},\\
\langle k^i_\perp\rangle_{\text{JM},\eta}&=-\langle k^i_\perp\rangle_{\text{JM},-\eta}.
\end{align}
It follows that
\begin{equation}\label{potPT}
\langle k^i_\perp\rangle_\text{Ji}=0\qquad\Rightarrow\qquad \langle k^i_\perp\rangle_{\text{pot},\eta}=\langle k^i_\perp\rangle_{\text{JM},\eta}=-\langle k^i_\perp\rangle_{\text{pot},-\eta}.
\end{equation}
In high-energy scatterings, the LF Wilson lines encode (leading-twist) initial and final-state interactions (ISI/FSI)~\cite{Ji:2002aa}. The choice of a particular boundary condition determines therefore the inclusion of these interactions in the definition of the canonical momentum. Except for $\eta=0$, ISI and FSI are included in an asymmetric way and allow for a nonvanishing average quark TM. Since time-reversal symmetry plays an important role, it is useful to split $A^\mu_{\text{phys},\eta}$ into \textsf{PT}-even and odd\footnote{In the literature one also often finds the terminology (naive) \textsf{T}-even and odd contributions.} contributions~\cite{Boer:2003cm}
\begin{equation}
A^\mu_{\text{phys},\eta}(x)=A^\mu_{\text{phys},e}(x)+\eta A^\mu_{\text{phys},o}(x)
\end{equation}
with
\begin{align}
A^\mu_{\text{phys},e}(x)&=\frac{1}{2}\left[A^\mu_{\text{phys},+}(x)+A^\mu_{\text{phys},-}(x)\right]\nn\\
&=-\frac{1}{2}\int_{-\infty^-}^{+\infty^-}\ud y^-\,\varepsilon(y^-)\,\mathcal W_\text{LF}(x^-,y^-;\tilde{\uvec x})\,F^{+\mu}(y^-,\tilde{\uvec x})\,\mathcal W_\text{LF}(y^-,x^-;\tilde{\uvec x}),\\
A^\mu_{\text{phys},o}(x)&=\frac{1}{2}\left[A^\mu_{\text{phys},+}(x)-A^\mu_{\text{phys},-}(x)\right]\nn\\
&=-\frac{1}{2}\int_{-\infty^-}^{+\infty^-}\ud y^-\,\mathcal W_\text{LF}(x^-,y^-;\tilde{\uvec x})\,F^{+\mu}(y^-,\tilde{\uvec x})\,\mathcal W_\text{LF}(y^-,x^-;\tilde{\uvec x}),
\end{align}
where $\varepsilon(y^-)$ is the sign function.\newline

In the following, we will compute directly the potential TM of quarks in the perturbative scalar diquark model (SDM) at order $\mathcal O(\lambda^2e_qe_s)$ and then compare it with the result one obtains for the JM quark TM. In this model, only abelian gauge interactions are considered, as any non-abelian effects would appear at higher order in perturbation theory. Expressions for the TM are therefore the same upon the substitution $g\mapsto -e_q$.

\subsection{Potential transverse momentum of quarks}
\label{PotMomentum}

\begin{figure}[h]
    \centering
    \includegraphics[scale=0.18]{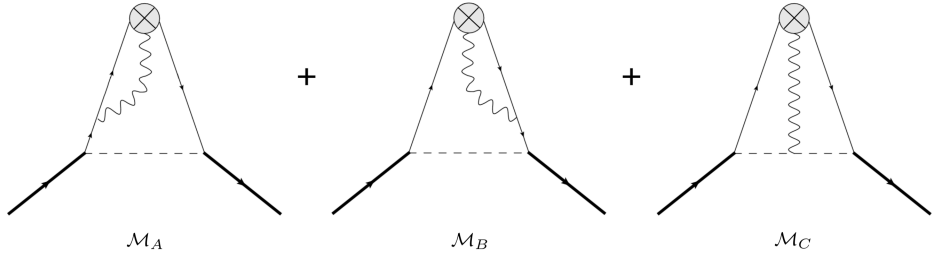}
    \caption{Diagrams contributing to the quark potential TM and AM at $\mathcal{O}(\lambda^2 e_q e_s)$ in perturbation theory. The dotted line represents the scalar diquark, the solid thin (thick) line represents the quark (nucleon), and the wavy line represents the photon. The crossed blob represents the insertion of an external operator.}
    \label{fig:M_I_II_III}
\end{figure}

In the SDM at order $\mathcal O(\lambda^2e_qe_s)$, the three diagrams shown in Fig.~\ref{fig:M_I_II_III} contribute a priori to the quark potential TM.
The first two diagrams represent the self-force since the Lorentz force felt by the quark finds its source in the quark itself. The last diagram represents the force exerted by the spectator diquark system. An explicit calculation outlined in~\ref{Sec_InII_PM} shows that
\begin{equation}
\mathcal M^i_{A+B}=0,
\end{equation}
This can be understood by the fact that this sum is even under \textsf{PT} whereas the potential TM is \textsf{PT}-odd. Also we can intuitively expect that there are no M\"unchhausen quarks, i.e. quarks that can self-accelerate.

All the quark potential TM comes therefore from the third diagram where the photon is attached to the spectator system. It contains both \textsf{PT}-even and odd contributions. A direct calculation confirms that the \textsf{PT}-even contribution vanishes. Using dimensional regularization with $D=4-2\epsilon$, we find for the \textsf{PT}-odd contribution
\begin{equation}\label{kpot}
\langle k^i_\perp\rangle_{\text{pot},\eta}=-\eta\,\epsilon^{ij}_\perp s^j_\perp\,\frac{\pi}{6}\,(3m_q+M)\, \frac{\mathcal N}{(4\pi\epsilon)^2}+\mathcal O(1/\epsilon)
\end{equation}
with $\mathcal N=\lambda^2e_qe_s/(4\pi)^2$ and $m_q$ the quark mass, see~\ref{Sec_III_PM} for details.

\subsection{Canonical transverse momentum of quarks}
\label{JM-TransvMomentum}

The JM quark TM can be expressed directly in terms of transverse-momentum dependent distributions (TMDs)~\cite{Mulders:1995dh,Diehl:2015uka}. The leading-twist vector TMD correlator is defined as
\begin{align}
&\Phi^{[\gamma^+]}_\eta(P,x,\uvec{k}_\perp,S)= \nn\\
&\qquad\qquad\frac{1}{2}\int\frac{\ud z^-\,\ud^2 z_\perp}{(2\pi)^3}\,e^{ik\cdot z} \langle P,S | \bar{\psi}(-\tfrac{z}{2})\gamma^+\mathcal W_\eta(-\tfrac{z}{2},\tfrac{z}{2})\psi(\tfrac{z}{2}) | P,S \rangle \Big|_{z^+=0},
\label{TMDcorrelator}
\end{align}
where $x=k^+/P^+$ is the fraction of longitudinal LF momentum carried by the quark. The LF Wilson line satisfies the differential equation
\begin{equation}\label{diffeq}
D^{\text{pure},\eta}_\mu\mathcal W_\eta(z,z_0)=0
\end{equation}
and ensures that the non-local correlator is gauge invariant. In practice, the boundary condition is determined by the scattering process at hand. In semi-inclusive deep-inelastic scattering (SIDIS) one has $\eta=+1$, whereas in Drell-Yan process one has $\eta=-1$. The Wilson lines $\mathcal W_\pm(-\tfrac{z}{2},\tfrac{z}{2})$ with finite transverse width look like staples and result from the infinite product of infinitesimally thin staples all with the same orientation. When $\eta\neq\pm 1$, the Wilson line is flaky and consists in infinitely many zigzags between $+\infty^-$ and $-\infty^-$. No physical process so far has been identified for this case.

Since the pure-gauge covariant derivative satisfies Eq.~\eqref{diffeq}, it follows after some algebra that~\cite{Boer:2003cm,Meissner:2007rx}
\begin{equation}
\langle k^i_\perp\rangle_{\text{JM},\eta}=\int\ud x\,\ud^2k_\perp\,k^i_\perp\,\Phi^{[\gamma^+]}_\eta(P,x,\uvec{k}_\perp,S).
\end{equation}
Using the generic parametrization of the correlator~\eqref{TMDcorrelator} in terms of TMD parton distribution functions\footnote{The $\eta$-dependence is often incorporated into the Sivers function $f^{\perp,\eta}_{1T}$ which then changes sign under \textsf{PT}~\cite{Collins:2002kn}.}
\begin{equation}
\Phi^{[\gamma^+]}_\eta(P,x,\uvec{k}_\perp,S)= f_1(x,\uvec k^2_\perp)-\eta\,\frac{\epsilon^{ij}_\perp  k^i_\perp S^j_\perp}{M}\,f^{\perp }_{1T}(x,\uvec k_\perp^2),
\end{equation}
one arrives at the expression for the so-called Sivers shift~\cite{Boer:2003cm,Meissner:2007rx}
\begin{equation}\label{Siversshift}
\langle k^i_\perp\rangle_{\text{JM},\eta}=-\eta\,\epsilon^{ij}_\perp s^j_\perp\int\ud x\,\ud^2k_\perp\,\frac{\uvec k^2_\perp}{2M}\,f^{\perp}_{1T}(x,\uvec k^2_\perp),
\end{equation}
where under the integral one has replaced $k^i_\perp k^l_\perp$ by $\delta^{il}_\perp\,\uvec k_\perp^2/2$. The function $f^{\perp}_{1T}$ has originally been introduced by Sivers~\cite{Sivers:1989cc} to explain large left-right asymmetries observed in pion-nucleus collisions~\cite{Antille:1980th}. The Sivers asymmetry has now been experimentally observed in both SIDIS~\cite{Airapetian:2004tw,Alexakhin:2005iw,Alekseev:2010rw} and Drell-Yan~\cite{Aghasyan:2017jop} processes, see also~\cite{Boer:2015vso} for a review on the gluonic counterpart.

The quark Sivers function within the SDM has been computed in~\cite{Ji:2002aa} with advanced boundary condition
\begin{equation} \label{JiGoekeSivers}
f^{\perp q}_{1T}(x,\uvec{k}^2_\perp)
=
\frac{\lambda^2 e_q e_s (1-x)}{4(2\pi)^4}\,
\frac{M(m_q + xM)}{\uvec k^2_\perp[\uvec{k}^{2}_{\perp}+\Lambda^2 (x)]}\,
\ln\! \left(
\frac{\uvec{k}^{2}_{\perp}+\Lambda^2 (x)}{\Lambda^2 (x)}
\right),
\end{equation}
where $\Lambda^2(x)=xm^2_s+(1-x)m^2_q-x(1-x)M^2$ with $m_s$ the scalar diquark mass. Inserting this expression into Eq.~\eqref{Siversshift} we obtain using dimensional regularization with $d\equiv D-2=2-2\epsilon$ for the transverse-momentum integral
\begin{equation}
\langle k^i_\perp\rangle_{\text{JM},+1}
\underset{\text{naive}}{=} -\eta\,\epsilon^{ij}_\perp s^j_\perp\,\frac{\pi}{3}\,(3m_q+M)\, \frac{\mathcal N}{(4\pi\epsilon)^2}+\mathcal O(1/\epsilon), \label{k_average}
\end{equation}
i.e. \emph{twice} $\langle k^i_\perp\rangle_{\text{pot},+1}$, in flagrant conflict with the expectation from \textsf{PT} symmetry~\eqref{potPT}! The solution to this conundrum is the realization that the expression in Eq.~\eqref{JiGoekeSivers} is not the appropriate one in dimensional regularization. In the derivation of the Sivers function, a transverse loop integral 
\begin{equation}
I_{d}=\int \frac{\ud^{d} l_\perp}{ (2\pi)^{d}}
\,\frac{\uvec{l}_\perp\cdot\uvec k_\perp}
{\uvec{l}^{2}_\perp[(\uvec{k}^{2}_{\perp}+\uvec{l}^{2}_{\perp})+\Lambda^2 (x)]}
\end{equation}
has been performed in $d=2$ dimensions, leading to the logarithmic factor~\cite{Meissner:2007rx}
\begin{equation}\label{MMGid}
I_2= -
\frac{1}{4\pi}\,
\ln \!\left(
\frac{\uvec{k}^{2}_{\perp}+\Lambda^2 (x)}{\Lambda^2 (x)}
\right).
\end{equation}
Even if this integral is finite in $d=2$ dimensions for finite $\uvec k^2_\perp$, we are actually interested in the leading divergence and hence in the large-$\uvec k^2_\perp$ region, which interferes with the $\epsilon$-expansion and must be treated with care~\cite{Davydychev:1993pg,Weinberg:1959nj,Smirnov:1990rz}. In dimensional regularization we need to keep $d\equiv D-2=2-2\epsilon_l$, leading to
\begin{equation}
I_{2-2\epsilon_l}=-\frac{\uvec k^2_\perp}{(4\pi)^{1-\epsilon_l}}\,\Gamma(1+\epsilon_l)\int_0^1\ud y\,y^{-\epsilon_l}\left[(1-y)\uvec k^2_\perp+\Lambda^2(x)\right]^{-1-\epsilon_l}.
\end{equation}
Proceeding with this expression instead of~\eqref{MMGid} and integrating over $\uvec k_\perp$ in $d=2-2\epsilon$ dimensions, we find\footnote{In the integral over $\uvec k_\perp$, we can replace the ratio $\uvec k^2_\perp/[\uvec k^2_\perp+\Lambda^2(x)]$ by $1$ since we are only interested in the leading divergence.}
\begin{equation}
\langle k^i_\perp\rangle_{\text{JM},+1}
= -\eta\,\epsilon^{ij}_\perp s^j_\perp\,\frac{\pi}{3}\,(3m_q+M)\, \frac{\mathcal N}{(4\pi)^2\epsilon(\epsilon+\epsilon_l)}+\mathcal O(1/\epsilon).
\end{equation}
The naive result~\eqref{k_average} corresponds to $\epsilon_l\to 0$, while the correct result $\langle k^i_\perp\rangle_{\text{JM},+1}=\langle k^i_\perp\rangle_{\text{pot},+1}$ is obtained when $\epsilon_l\to\epsilon$. The same conclusion is reached using Weinberg's theorem to extract the correct large-$\uvec k^2_\perp$ behavior of $I_d$. More generally, this is in line with the observation from Refs.~\cite{Kang:2010hg,Qiu:2020oqr} that Eq.~\eqref{Siversshift} holds only if the \emph{same} regularization and renormalization procedures are applied on both sides of the identity. Note however that the explicit check of the Burkardt sum rule of Ref.~\cite{Goeke:2006ef} is not impacted since the same subtlety applies equally to the gluon contribution.

\section{Potential angular momentum of quarks}
\label{PotOAM}

Similarly to the quark potential TM, we can now consider the quark potential longitudinal AM~\cite{Wakamatsu:2010qj,Hatta:2011ku} at equal LF time $r^+=0$
\begin{align}
\langle L_z\rangle_{\text{pot},\eta}&=\lim_{\Delta\to 0}\frac{\langle p',S'|\int\ud^4 r\,\delta(r^+)\,(-g)\overline\psi(r)\gamma^+[\uvec r_\perp\times\uvec A_{\text{phys},\eta,\perp}(r)]_z\psi(r)|p,S\rangle}{\langle P,S|P,S\rangle}\nn\\
&=\frac{-g\,\epsilon^{ij}_\perp}{2P^+}\left[-i\nabla^i_{\Delta_\perp}\langle p',S'|\overline\psi(0)\gamma^+A^j_{\text{phys},\eta,\perp}(0)\psi(0)|p,S\rangle\right]_{\Delta=0},\label{defLpot}
\end{align}
where $p=P-\tfrac{\Delta}{2}$ and $p'=P+\tfrac{\Delta}{2}$. The covariant LF polarization vectors $S$ and $S'$ refer to the same rest-frame spin vector $\uvec s$. In order to deal properly with the explicit factor of $\uvec r_\perp$ in the operator, the matrix element should first be considered with $p'\neq p$, and the forward limit $\Delta\equiv p'-p\to 0$ be taken at the end~\cite{Bakker:2004ib,Lorce:2011ni}. This potential AM represents the difference between the JM and Ji definitions of quark OAM, obtained by replacing $-g A^j_{\text{phys},\eta,\perp}$ in Eq.~\eqref{defLpot} with $\frac{i}{2}\overset{\leftrightarrow}{D}\!\!\!\!\!\phantom{D}^j_{\text{pure},\eta,\perp}$ and $\frac{i}{2}\overset{\leftrightarrow}{D}\!\!\!\!\!\phantom{D}^j_\perp$, respectively. Discrete space-time symmetries indicate that both JM and Ji OAM are \textsf{PT}-even, and so is the potential AM
\begin{equation}
\langle L_z\rangle_{\text{pot},\eta}=\langle L_z\rangle_{\text{pot},-\eta}.
\end{equation}
This means that the boundary condition has no effects on the value of the longitudinal OAM~\cite{Hatta:2011ku,Lorce:2012ce} and we can drop the reference to the parameter $\eta$.\newline

At one-loop level in the SDM, it has been shown that JM and Ji notions of OAM computed independently do coincide~\cite{Burkardt:2008ua,Lorce:2017wkb}, implying therefore a vanishing potential AM. This is of course expected since at that order no gauge field is involved. Perhaps more surprisingly, for an electron in QED where the gauge field appears already at the one-loop level, one obtains also a vanishing potential AM at that order~\cite{Ji:2015sio}. So the presence of a gauge field is not a sufficient condition for inducing a difference between JM and Ji OAM. As already mentioned in the introduction, calculations on the lattice indicate a clear non-vanishing potential AM~\cite{Engelhardt:2017miy,Engelhardt:2020qtg}. This motivates us to investigate the potential AM at the two-loop level within the SDM.

To the best of our knowledge, JM and Ji notions of OAM have never been calculated in the SDM at order $\mathcal O(\lambda^2 e_qe_s)$. Since we are in fact interested in their difference, we compute directly the quark potential AM from the diagrams shown in Fig.~\ref{fig:M_I_II_III}. Contrary to the potential TM, the potential AM is \textsf{PT}-even and hence can a priori receive a non-vanishing contribution from both mechanisms $\mathcal M_{A+B}$ and $\mathcal M_C$. An explicit calculation outlined in~\ref{Sec_InII} shows that at best
\begin{equation}
\left[-i\epsilon^{ij}_\perp\nabla^i_\perp\mathcal M^j_{A+B}(\Delta)\right]_{\Delta=0}=\mathcal O(1/\epsilon).
\end{equation}
The absence of $1/\epsilon^2$-divergences can intuitively be understood by the fact that a quark should not be able to exert a self-torque and hence change its own OAM, which would lead to a violation of AM conservation\footnote{Since we are looking at the $\gamma^+$ component, the quark LF helicity is preserved at the level of the operator insertion and there is no spin-orbit conversion.}. All the $1/\epsilon^2$-contributions to the quark potential AM come therefore from the third diagram where the photon is attached to the spectator system. Using dimensional regularization, we actually find that all $1/\epsilon^2$-divergences cancel out, so that we have at best
\begin{equation}
\langle L_z\rangle_\text{pot}= \mathcal O(1/\epsilon),
\label{final_result}    
\end{equation}
see~\ref{Sec_III} for more details. This is somewhat surprizing at first sight since ISI/FSI, represented at this order by the photon exchange with the spectator system, has already been shown to exert a Lorentz force on the struck quark. On a second thought, for a nonvanishing potential AM we need a Lorentz torque defined relative to the center of inertia of the system~\cite{Burkardt:2002ks,Lorce:2018zpf}. In the SDM, the nucleon is described as a two-body system. The positions of the quark, the diquark, and the center of inertia (or $P^+$) all lie on the same line, preventing therefore classically any Lorentz torque.

\section{Summary and Outlook}
\label{Conclusions}
 
The potential transverse momentum and potential longitudinal angular momentum, describing the difference between the corresponding Jaffe-Manohar and Ji definitions of linear and angular momenta, have been computed for the first time at order $\mathcal{O}(\lambda^2 e_q e_s)$ in the scalar diquark model. At that order, two mechanisms are possible: one where a gauge boson is exchanged between the struck quark and the spectator system, and one where the struck quark exchanges a gauge boson with itself. As expected by conservation of linear and angular momentum, the latter mechanism does not contribute (at least to the $1/\epsilon^2$-divergence). The former mechanism is responsible for the Sivers shift and represents the accumulated exchange of momentum between the struck quark and the spectator system in a high-energy process. Our explicit model calculation confirms that the potential transverse momentum coincides with the average Jaffe-Manohar (or canonical) transverse momentum of the quark, as expected by parity and time-reversal symmetries. For the potential angular momentum, we found that all $1/\epsilon^2$-divergent terms cancel out in the scalar diquark model at two loops. We suggested that such a result could be the reflection of the fact that in a two-body description of a system, there cannot be classically any Lorentz torque exerted by the spectator on the struck constituent despite the presence of a Lorentz force. A confirmation of these observations in QED or the quark-target model would be very interesting but is left for future works.

\begin{acknowledgements}
We are grateful to A. Metz and U. Reinosa for helping us understand the conundrum with the calculation of the Sivers shift. D. A. Amor-Quiroz acknowledges financial support from Consejo Nacional de Ciencia y Tecnolog\'ia (CONACyT) postdoctoral fellowship number 711226 (CVU 449539) and a one-year postdoctoral fellowship provided by Ecole Polytechnique. M. Burkardt was supported by the DOE under grant number DE-FG03-95ER40965 and C. Lorc\'e by the Agence Nationale de la Recherche under the Projects No. ANR-18-ERC1-0002 and No. ANR-16-CE31-0019.
\end{acknowledgements}

\appendix

\section{Scalar diquark model}
\label{AppSDM}

\begin{figure}[htb]
    \centering
    \includegraphics[scale=0.22]{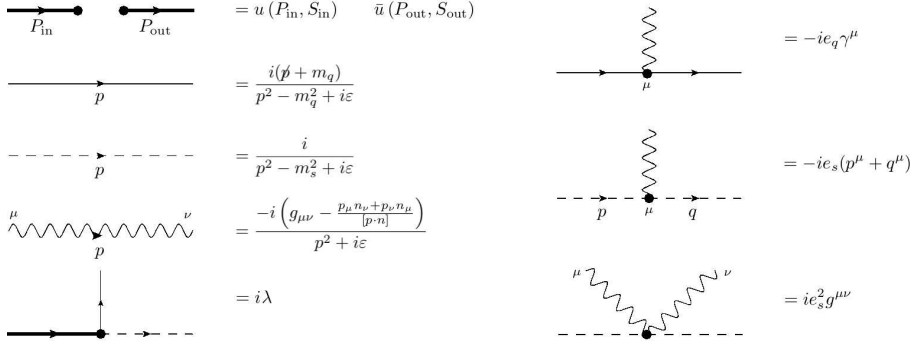}
    \caption{Feynman rules for the scalar diquark model with a pointlike quark-nucleon-diquark coupling concerning the lagrangian from Eq.~\eqref{Lagrangian}. The photon propagator is considered in the light-front gauge $A^+\equiv A\cdot n=0$ with $n=(1,0,0,-1)/\sqrt{2}$. The prescription for the $\frac{1}{[p\cdot n]}$ pole depends on the choice of the boundary condition~\cite{Belitsky:2002sm}. Concerning the $p_\nu n_\mu$ term one has $\frac{1}{p\cdot n-i\varepsilon}$ for advanced, $\frac{1}{p\cdot n+i\varepsilon}$ for retarded, and $\frac{1}{2}[\frac{1}{p\cdot n-i\varepsilon}+\frac{1}{p\cdot n+i\varepsilon}]$ for antisymmetric boundary condition. The other term is related by hermiticity.}
    \label{fig:Rules}
\end{figure}

The lagrangian for the SDM is
\begin{align}
\mathcal{L}_{\text{SDM}} 
&=
-\frac{1}{4}\,F^{\mu\nu}F_{\mu\nu}+
\bar{\Psi}_{N} \left( i\fsl{\partial}-M \right) \Psi_{N}
+
\bar{\psi} \left( i\fsl{\partial}-m_q \right) \psi
+
\partial_\mu\phi^* \partial^\mu\phi 
- 
m^2_s \phi^* \phi
\nn\\
& +
\lambda  \left( \bar{\psi}\Psi_{N}\phi^* + \bar{\Psi}_{N}\psi\phi \right) 
-
e_q \bar{\psi} \fsl{A} \psi
-
i e_s \big( \phi^* \overset{\leftrightarrow}{\partial^\mu} \phi \big) A_\mu
+
e^2_s A_\mu A^\mu \phi^* \phi ~.
\quad
\label{Lagrangian}
\end{align}
In this expression, $\psi$ represents the quark field with mass $m_q$, $\phi$ denotes the charged scalar diquark field with mass $m_s$, $\Psi_N$ is the neutral nucleon field with mass $M$, and $A^\mu$ is a gauge field that could represent either photons or gluons. The stability condition for the target nucleon is $M<m_q+m_s$. Furthermore, the photon-quark and photon-diquark couplings are given by $e_q$ and $e_s$ respectively, while $\lambda$ is the coupling constant of the point-like scalar quark-nucleon-diquark vertex. The corresponding Feynman rules are summarized in fig.~\ref{fig:Rules}.

The nucleon field considered in the present model has no charge emulating either a neutron in QED, or the fact that in QCD a hadron is color neutral. Such condition simplifies the calculations as no gauge field can couple to the target. Moreover, it implies that the photon-quark and photon-diquark coupling constants are equal but with opposite sign, for instance $e_q=-e_s$ for QED. For the corresponding QCD generalization of our results, it suffices to perform the replacement $e_qe_s \mapsto 4\pi C_F \alpha_s$~\cite{Brodsky:2002cx} since non-abelian aspects appear only at higher order.

For convenience, we will consider the potential TM and AM with $\eta=+1$ in the corresponding natural gauge, namely the LF gauge with advanced boundary condition. In this case $A^\mu_\text{phys}(x)$ simply reduces to $A^\mu(x)$.

\section{Quark potential linear momentum in the SDM}

In the SDM at order $\mathcal O(\lambda^2 e_qe_s)$, the quark TM arises from the three diagrams shown in Fig.~\ref{fig:M_I_II_III}. The first two, denoted $\mathcal M_A$ and $\mathcal M_B$, correspond to the mechanism where the probed photon is attached to a quark line. The third diagram, denoted $\mathcal M_C$, corresponds to the mechanism where the probed photon is attached to the spectator diquark. In the following, we first check explicitly that the sum of the contributions from the first two diagrams vanishes. We then proceed with the calculation of the third contribution.

\subsection{Diagrams $\mathcal M_A$ and $\mathcal M_B$}
\label{Sec_InII_PM}

\begin{figure}[h]
    \centering
    \includegraphics[scale=0.26]{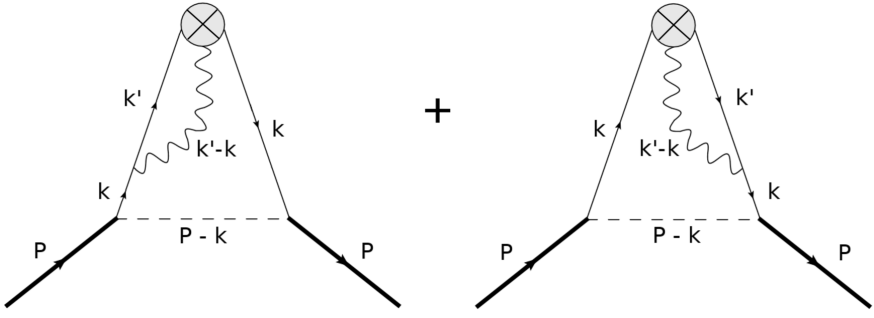}
    \caption{Sum of the diagrams with a diquark acting as a spectator system while a photon is attached to either the incoming (left) or the outgoing (right) quark.}
    \label{fig:M_I_IIPTM}
\end{figure}
The amplitudes associated with the Feynman diagrams from Fig.~\ref{fig:M_I_IIPTM} read in $D$ close to four dimensions
\begin{align}
\mathcal{M}^{i}_A&=
\frac{\lambda^2e^2_q}{2P^+}
\int \frac{\ud^D k}{(2\pi)^D}
\int \frac{\ud^D k'}{(2\pi)^D}
\left( g^{i\nu}-\frac{(k'-k)^i_\perp n^\nu}{(k'-k)^++i\varepsilon}\right)\nn\\
&\quad\times\frac{N^+_{\phantom{+}\nu}(k,k')}{D(k'-k,0)\,D(P-k,m_s)\,[D(k,m_q)]^2\,D(k',m_q)},\\
\mathcal{M}^{i}_B&=\frac{\lambda^2e^2_q}{2P^+}
\int \frac{\ud^D k}{(2\pi)^D}
\int \frac{\ud^D k'}{(2\pi)^D}
\left( g^{i\nu}-\frac{(k'-k)^i_\perp n^\nu}{(k'-k)^+-i\varepsilon}\right)\nn\\
&\quad\times\frac{N^{\phantom{\nu}+}_\nu(k,k')}{D(k'-k,0)\,D(P-k,m_s)\,[D(k,m_q)]^2\,D(k',m_q)},\label{MABpotTM}
\end{align}
where
\begin{align}
N^{\mu\nu}(k,k')&=\bar{u}(P,S)(\fsl{k}+m_q ) \gamma^\mu(\fsl{k}'+m_q )\gamma^\nu(\fsl{k}+m_q )u(P,S),\label{Nstrucdef}\\
D(k,m)&=k^2-m^2+i\varepsilon.
\end{align}
Working out the numerator we get
\begin{align}
&N^{+i}(k,k')+N^{i+}(k,k')- 2N^{++}(k,k')\,\text{p.v.}\!\left[\frac{(k'-k)^i_\perp}{(k'-k)^+}\right]\nn\\
&\qquad=
-4(k^2-m_q^2)\left[k'^i_\perp P^+ + k'^+P^i_\perp\right] +8(P\cdot k+m_q M) \left[k^+k'^i_\perp+k^i_\perp k'^+\right]\nn\\
&\qquad\quad + \text{p.v.}\!\left[\frac{8(k'-k)^i_\perp k'^+ }{(k'-k)^+}\right]\left[( k^2-m_q^2) P^+ - 2k^+  (P\cdot k + m_q M)\right],
\label{AmpInIIPM}
\end{align}
where p.v. stands for Cauchy's principal value. Note that in the symmetric LF frame $\uvec P_\perp=\uvec 0_\perp$, there is no transverse vector in Eq.~\eqref{MABpotTM} left after integration over $k$ and $k'$, and so
\begin{equation}
\mathcal{M}^{i}_{A+B}  = 0 .
\end{equation}
From a slightly different perspective, we know from \textsf{P} and \textsf{T} symmetries that $\mathcal M^i\propto \epsilon^{ij}_\perp S^j_\perp$. Since any possible dependence on the target polarization comes from $N^{\mu\nu}(k,k')$, its absence in Eq.~\eqref{AmpInIIPM} implies that the amplitude must vanish.

\subsection{Diagram $\mathcal M_C$}
\label{Sec_III_PM}

\begin{figure}[h]
    \centering
    \includegraphics[scale=0.26]{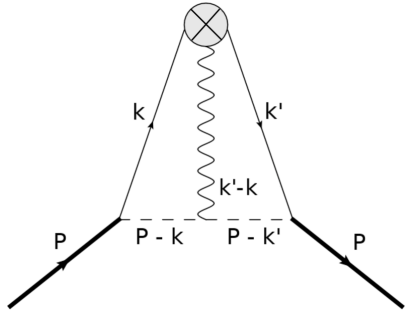}
    \caption{Diagram with a diquark acting as a spectator system while a photon is attached to it.}
    \label{fig:M_IIIPTM}
\end{figure}
The amplitude associated with the Feynman diagram shown in Fig.~\ref{fig:M_IIIPTM} reads in $D$ close to four dimensions
\begin{align}
\mathcal{M}^{i}_{C}& =\frac{\lambda^2 e_qe_s}{2P^+}\int \frac{\ud^D k}{(2\pi)^D}\int \frac{\ud^D k^\prime}{(2\pi)^D}\left( g^{i\nu}-\frac{(k'-k)^i_\perp n^\nu}{(k'-k)^+-i\varepsilon}\right)\nn\\
&\quad\times\frac{(2P-k'-k)_\nu\,\bar{u}(P,S)(\fsl{k}^\prime+m_q)\gamma^+(\fsl{k}+m_q )u(P,S)}{D(k'-k,0)\,D(P-k,m_s)\,D(P-k',m_s)\,D(k,m_q)\,D(k',m_q)}.\label{AmpIIIPM}
\end{align}
Working out the numerator we get ($\epsilon_{0123}=+1$)
\begin{align}
&\bar{u}(P,S)(\fsl{k}^\prime+m_q )  \gamma^+(\fsl{k}+m_q )u(P,S) \nn\\
&\qquad= 2\left[k^+(P\cdot k') + k'^+(P\cdot k) \right] -2P^+(k'\cdot k - m_q^2)  + 2m_q M(k'+k)^+  \nn\\
&\qquad\quad+ 2i\epsilon^{+\alpha\beta\lambda}S_\lambda \left[M k'_\alpha k_\beta +m_q (k'-k)_\alpha P_\beta\right].\label{Dirac_MIIIPM}
\end{align}
In the symmetric LF frame $\uvec P_\perp=\uvec 0_\perp$, the only term in Eq.~\eqref{Dirac_MIIIPM} surviving integration over $k$ and $k'$ is the one involving the target polarization. Since it is antisymmetric under $k'\leftrightarrow k$ whereas the combination of denominators in Eq.~\eqref{AmpIIIPM} is symmetric, we just need to keep the antisymmetric part of the gauge boson numerator $\left(g^{i\nu}-\frac{(k'-k)^i_\perp n^\nu}{(k'-k)^+-i\varepsilon}\right)\mapsto-(k'-k)^i_\perp n^\nu i\pi\,\delta(k'^+-k^+)$, also known as the gauge boson pole. We then find
\begin{align}
\mathcal{M}^{i}_{C}&=\epsilon_\perp^{ij}s^j_\perp\,\frac{\lambda^2e_qe_s}{P^+}\int\frac{\ud^Dk}{(2\pi)^D}\int\frac{\ud^Dk'}{(2\pi)^D}\,\pi\,\delta(k'^+-k^+)\nn\\
&\quad\times\frac{(P^+-k^+)(m_qP^++Mk^+)}{D(P-k,m_s)\,D(P-k',m_s)\,D(k,m_q)\,D(k',m_q)},
\end{align}
where we made the replacement $\frac{(k'-k)^i_\perp(k'-k)^j_\perp}{(k'-k)^2+i\varepsilon}\mapsto -\frac{\delta^{ij}_\perp}{2}$ valid\footnote{Strictly speaking, it should be $-\frac{\delta^{ij}_\perp}{D-2}$ but this does not make any difference for the leading divergence in $D=4-2\epsilon$.} under the integrals thanks to $\delta(k'^+-k^+)$. The integral over $k^-$ can be performed by closing the integration contour on the lower half of the complex $k^-$-plane
\begin{equation}
\int \frac{\ud k^{-}}{2\pi}
\frac{1}{D(P-k,m_s)\,D(k,m_q)}=\frac{i}{2P^+}\,\frac{1}{\uvec{k}^{2}_\perp + \Lambda^2(x,1)},
\end{equation}
and similarly for the integral over $k'^-$. We introduced for convenience the quark longitudinal LF momentum fractions $x=k^+/P^+$ and $x'=k'^+/P^+$, and the effective squared-mass function
\begin{align}
\Lambda^2(x,y)=yxm^2_s+(1-x)m^2_q-yx(1-x)M^2
\label{Lambda_def}
\end{align}
which is positive-definite for $x,y\in[0,1]$ owing to the stability condition $M<m_q+m_s$. Performing the integrals over the transverse momenta in $d=2-2\epsilon$ dimensions and over the longitudinal momenta, we find
\begin{align}
\mathcal{M}^{i}_{C}&=-\epsilon^{ij}_\perp s_\perp^j\,\pi\mathcal N\int\ud x\,(1-x)(m_q+xM)\left[\int\frac{\ud^dk_\perp}{(2\pi)^d}\,\frac{1}{\uvec k^2_\perp+\Lambda^2(x,1)}\right]^2\nn\\
&=-\epsilon^{ij}_\perp s^j_\perp\,\frac{\pi}{6}\,(3m_q+M)\, \frac{\mathcal N}{(4\pi\epsilon)^2}+\mathcal O(1/\epsilon)
\end{align}
with $\mathcal N=\lambda^2e_qe_s/(4\pi)^2$.

\section{Quark potential angular momentum in the SDM}

In the SDM at order $\mathcal O(\lambda^2 e_qe_s)$, the quark potential AM arises again from the three diagrams shown in Fig.~\ref{fig:M_I_II_III}. Contrary to the TM, the contribution from the sum of the first two diagrams, representing the self-torque, does not need to vanish due to \textsf{PT} symmetry. For convenience, we will work in the symmetric Drell-Yan frame defined by $\Delta^+=0$ and $\uvec P_\perp=\uvec 0_\perp$.

\subsection{Diagrams $\mathcal M_A$ and $\mathcal M_B$}
\label{Sec_InII}

\begin{figure}[h]
    \centering
    \includegraphics[scale=0.26]{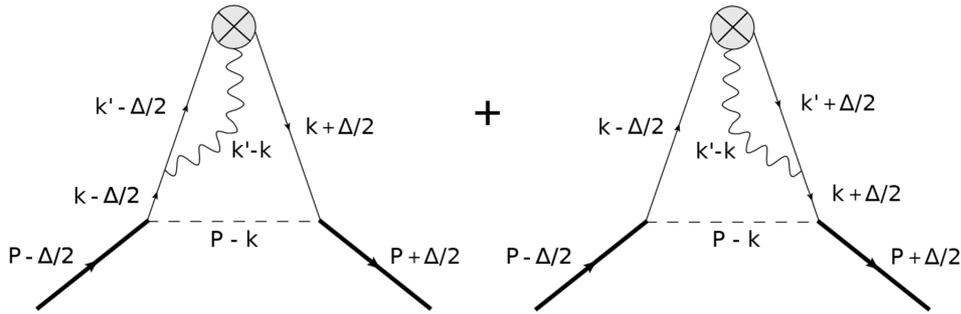}
    \caption{Sum of the diagrams with a diquark acting as a spectator system while a photon is attached to either the incoming (left) or the outgoing (right) quark.}
    \label{fig:M_I_II}
\end{figure}
The amplitudes associated with the Feynman diagrams from Fig.~\ref{fig:M_I_II} read in $D$ close to four dimensions
\begin{align}
\mathcal{M}^{j}_A(\Delta)&=
\frac{\lambda^2e^2_q}{2P^+}
\int \frac{\ud^D k}{(2\pi)^D}
\int \frac{\ud^D k'}{(2\pi)^D}
\left( g^{j\nu}-\frac{(k'-k)^j_\perp n^\nu}{(k'-k)^++i\varepsilon}\right)\frac{1}{D(k'-k,0)}\nn\\
&\quad\times\frac{N^+_{\phantom{+}\nu}(k,k'-\tfrac{\Delta}{2},\Delta)}{D(P-k,m_s)\,D(k+\tfrac{\Delta}{2},m_q)\,D(k-\tfrac{\Delta}{2},m_q)\,D(k'-\tfrac{\Delta}{2},m_q)},\label{AmpInIIPAM1}\\
\mathcal{M}^{j}_B(\Delta)&=\frac{\lambda^2e^2_q}{2P^+}
\int \frac{\ud^D k}{(2\pi)^D}
\int \frac{\ud^D k'}{(2\pi)^D}
\left( g^{j\nu}-\frac{(k'-k)^j_\perp n^\nu}{(k'-k)^+-i\varepsilon}\right)\frac{1}{D(k'-k,0)}\nn\\
&\quad\times\frac{N_\nu^{\phantom{\nu}+}(k,k'+\tfrac{\Delta}{2},\Delta)}{D(P-k,m_s)\,D(k+\tfrac{\Delta}{2},m_q)\,D(k-\tfrac{\Delta}{2},m_q)\,D(k'+\tfrac{\Delta}{2},m_q)},\label{AmpInIIPAM2}
\end{align}
where 
\begin{equation}
N^{\mu\nu}(k,k',\Delta)=\bar u'(\fsl{k}+\tfrac{\fsl{\Delta}}{2}+m_q ) \gamma^\mu(\fsl{k}'+m_q )\gamma^\nu(\fsl{k}-\tfrac{\fsl{\Delta}}{2}+m_q )u
\end{equation}
using the shorthand notation for the spinors $\bar{u}' \equiv \bar{u}(P+\tfrac{\Delta}{2},S^\prime)$ and $ u\equiv u(P-\tfrac{\Delta}{2},S)$, with $S'$ and $S$ the covariant polarization vectors referring to the same rest-frame polarization $\uvec s$. Since we are interested in the potential AM, we need only to keep track of the terms linear in the momentum transfer, see Eq.~\eqref{defLpot}. Moreover, we know from discrete symmetries that the potential (longitudinal) AM is proportional to the target longitudinal polarization. Projecting out the contributions depending on the latter, we find using Dirac bilinear identities from~\cite{Lorce:2017isp}
\begin{align}
&\frac{1}{2}\sum_{s_z=\pm 1}s_z\left[N^{+j}(k,k')-N^{j+}(k,k')\right] \nn\\
&\qquad\qquad=-4\left[2(P\cdot k+m_qM)k^+-(k^2-m_q^2)P^+\right]i\epsilon^{jl}_\perp k'^l_\perp\nn\\
&\qquad\qquad\quad+8\left[(P\cdot k'+m_qM)k^+-(k'\cdot k-m_q^2)P^+\right]i\epsilon^{jl}_\perp k^l_\perp\nn\\
&\qquad\qquad\quad-8P^+k^j_\perp i\epsilon^{lm}_\perp k'^l_\perp k^m_\perp\\
&\frac{1}{2}\sum_{s_z=\pm 1}s_z\, N^{++}(k,k'\pm\tfrac{\Delta}{2},\Delta) \nn\\
&\qquad\qquad=-4k'^+(P-k)^+\,i\epsilon^{lm}_\perp k^l_\perp \Delta^m_\perp +\mathcal O(\Delta^2),\\
&\frac{1}{2}\sum_{s_z=\pm 1}s_z\left[N^{+j}(k,k'-\tfrac{\Delta}{2},\Delta)+N^{j+}(k,k'+\tfrac{\Delta}{2},\Delta)\right] \nn\\
&\qquad\qquad=-2k'^+\left[2(P\cdot k+m_qM)-(k^2-m_q^2)\right]i\epsilon^{jl}_\perp \Delta^l_\perp\nn\\
&\qquad\qquad\quad+2P^+\left[(m_q+\tfrac{k^+}{P^+}\,M)^2-\uvec k^2_\perp\right]i\epsilon^{jl}_\perp \Delta^l_\perp\nn\\
&\qquad\qquad\quad+4\left[k^j_\perp k'^+-(P-k)^+k'^j_\perp\right]i\epsilon^{lm}_\perp k^l_\perp \Delta^m_\perp +\mathcal O(\Delta^2).
\end{align}
Studying the position of the poles in Eqs.~\eqref{AmpInIIPAM1} and~\eqref{AmpInIIPAM2} provides the support properties $P^+>k^+>k'^+>0$. Closing the contours in the upper half of the complex planes, we can write
\begin{align}
&\int\frac{\ud k^-}{2\pi}\int\frac{\ud k'^-}{2\pi}\frac{1}{D(k'-k,0)\,D(P-k,m_s)\,[D(k,m_q)]^a\,[D(k',m_q)]^b}\nn\\
&\quad=\frac{-i}{2x(1-y)P^+}\,\frac{-i}{2(1-x)P^+}\,\frac{[-(1-x)]^{a+b}}{[D(\uvec k^2_\perp)]^a\,[D(\uvec k^2_\perp,\uvec q^2_\perp)]^b}
\end{align}
with $a,b>0$ and
\begin{align}
D(\uvec k^2_\perp)&=\uvec k^2_\perp+\Lambda^2(x,1),\\
D(\uvec k^2_\perp,\uvec q^2_\perp)&=y\uvec k^2_\perp+\Lambda^2(x,y)+\frac{1-x}{1-y}\uvec q_\perp^2,
\end{align}
where we introduced for convenience the shifted transverse momentum $\uvec q_\perp=\uvec k'_\perp-y\uvec k_\perp$.

After some algebra we find that the contribution from diagrams $\mathcal M_A$ and $\mathcal M_B$ to the quark potential AM is given by
\begin{equation}
\left[-i\epsilon^{ij}_\perp\nabla^i_\perp\mathcal M^j_{A+B}(\Delta)\right]_{\Delta=0}=-s_z\,\frac{\lambda^2e_q^2}{2(2\pi)^2}\int_0^1\ud x\,\ud y\,I_{A+B}(x,y),
\end{equation}
where
\begin{align}
I_{A+B}(x,y)&=\frac{(1-x)^2}{(1-y)}\int\frac{\ud^d k_\perp}{(2\pi)^d}\,\frac{\ud^d q_\perp}{(2\pi)^d}\left[-\frac{\uvec k^2_\perp-(1-y)(m_q+xM)^2}{[D(\uvec k^2_\perp)]^2\,D(\uvec k^2_\perp,\uvec q_\perp^2)}\right.\nn\\
&\quad+\frac{(1-x)\left\{[\uvec k^2_\perp-(m_q+xM)^2]\uvec q^2_\perp+2m_q(m_q+xM)y(1-y)\uvec k^2_\perp\right\}}{[D(\uvec k^2_\perp)]^2\,[D(\uvec k^2_\perp,\uvec q_\perp^2)]^2}\nn\\
&\quad\left.+\frac{y^2\uvec k^2_\perp}{D(\uvec k^2_\perp)\,[D(\uvec k^2_\perp,\uvec q_\perp^2)]^2}+\frac{y}{D(\uvec k^2_\perp)\,D(\uvec k^2_\perp,\uvec q_\perp^2)}\right].
\end{align}
Performing the integrals over the transverse momenta in $d=2-2\epsilon$, we find that the $1/\epsilon^2$-divergences cancel out, and so at best
\begin{equation}
\left[-i\epsilon^{ij}_\perp\nabla^i_\perp\mathcal M^j_{A+B}(\Delta)\right]_{\Delta=0}=\mathcal O(1/\epsilon).    
\end{equation}

\subsection{Diagram $\mathcal M_C$}
\label{Sec_III}

\begin{figure}[h]
    \centering
    \includegraphics[scale=0.26]{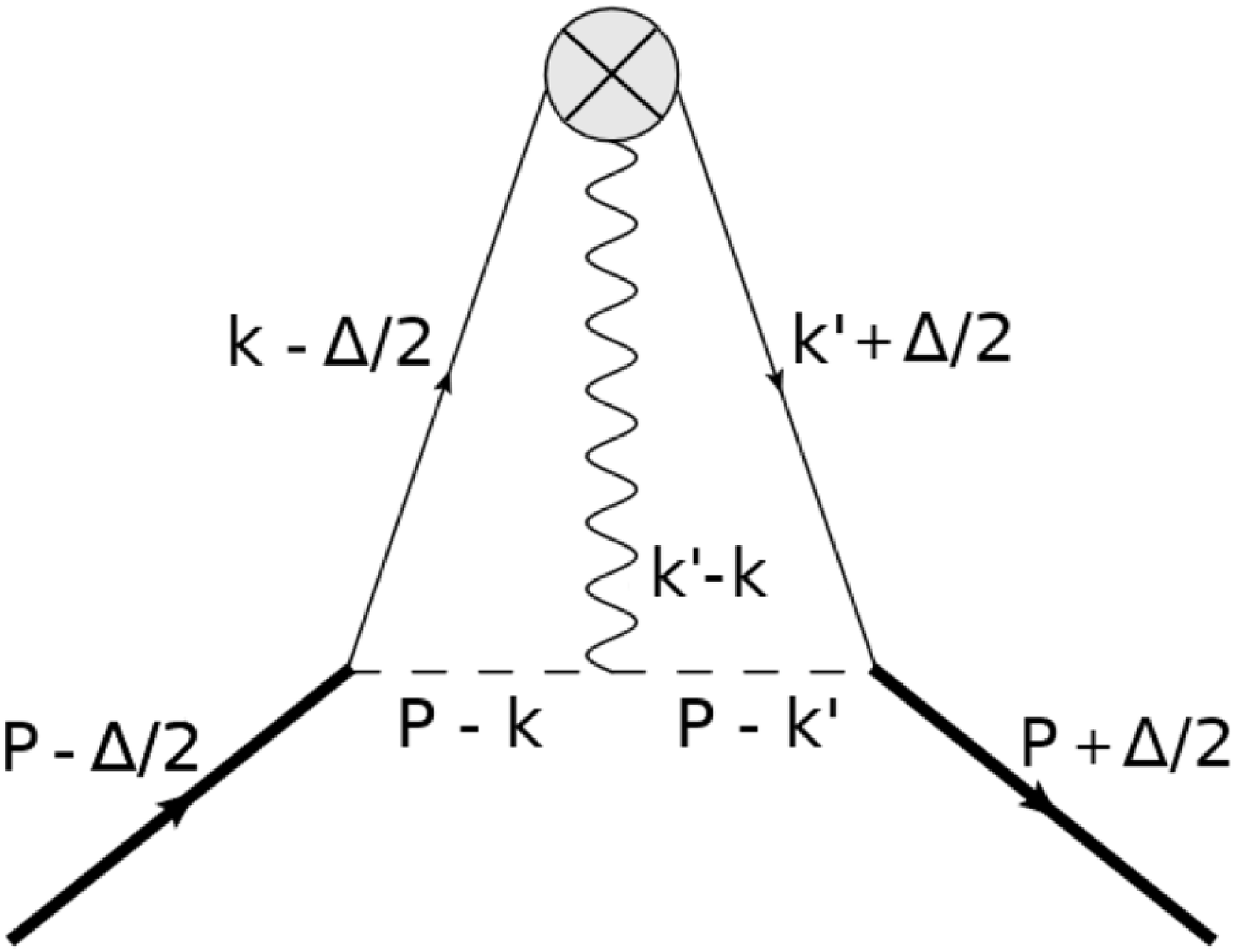}
    \caption{Diagram with the scalar diquark acting as the spectator system while a photon is attached to it.}
    \label{fig:M_III}
\end{figure}
The amplitude associated with the Feynman diagram shown in Fig.~\ref{fig:M_III} reads in $D$ close to four dimensions
\begin{align}
\mathcal{M}^{j}_{C}(\Delta)& =\frac{\lambda^2 e_qe_s}{2P^+}\int \frac{\ud^D k}{(2\pi)^D}\int \frac{\ud^D k^\prime}{(2\pi)^D}\left( g^{j\nu}-\frac{(k'-k)^j_\perp n^\nu}{(k'-k)^+-i\varepsilon}\right)\frac{1}{D(k'-k,0)}\nn\\
&\quad\times\frac{(2P-k'-k)_\nu\,\bar{u}'(\fsl{k}^\prime+\frac{\fsl{\Delta}}{2}+m_q)\gamma^+(\fsl{k}-\frac{\fsl{\Delta}}{2}+m_q )u}{D(P-k,m_s)\,D(P-k',m_s)\,D(k-\frac{\Delta}{2},m_q)\,D(k'+\frac{\Delta}{2},m_q)}.\label{AmpIIIAM}
\end{align}
Proceeding as in~\ref{Sec_InII}, we first focus on the Dirac structure in the numerator and project out the terms proportional to the nucleon longitudinal polarization $s_z$ up to linear order in $\Delta$, namely
\begin{align}
&\frac{1}{2}\sum_{s_z=\pm 1}s_z\,\bar{u}^\prime (\fsl{k}^\prime +\tfrac{\fsl{\Delta}}{2}+m_q )\gamma^+(\fsl{k}-\tfrac{\fsl{\Delta}}{2}+m_q)u\nn\\
&\quad=2 P^+ i\epsilon^{lm}_\perp k'^l_\perp k^m_\perp-\left[(P-k)^+ k'^l_\perp+(P- k')^+ k^l_\perp\right]i\epsilon^{lm}_\perp \Delta^m_\perp+ \mathcal{O}(\Delta^2).
\end{align}
Expanding the denominators
\begin{align}
&\frac{1}{D(k-\tfrac{\Delta}{2},m_q)\,D(k'+\tfrac{\Delta}{2},m_q)}\nn\\
&\qquad=\frac{1}{D(k,m_q)\,D(k',m_q)}\left[1+\frac{k\cdot\Delta}{D(k,m_q)}-\frac{k'\cdot\Delta}{D(k',m_q)}\right]+\mathcal O(\Delta^2)
\end{align}
we can use the symmetry under $k\leftrightarrow k'$ to restrict ourselves to the region $P^+>k^+>k'^+>0$, and close the $k^-$ ($k'^-$) contour in the upper (lower) half of the complex plane.

After some algebra we find that the contribution from diagram $\mathcal M_C$ to the quark potential AM is given by
\begin{equation}
\left[-i\epsilon^{ij}_\perp\nabla^i_\perp\mathcal M^j_{C}(\Delta)\right]_{\Delta=0}=-s_z\,\frac{\lambda^2e_q^2}{2(2\pi)^2}\int_0^1\ud x\,\ud x'\,I_{C}(x,x'),
\end{equation}
where
\begin{align}
I_{C}(x,x')&=\frac{\theta(x-x')}{(x-x')^2}\int\frac{\ud^d k_\perp}{(2\pi)^d}\,\frac{\ud^d k'_\perp}{(2\pi)^d}\frac{x'(1-x)}{D(\uvec k^2_\perp)D(\uvec k'^2_\perp)D(\uvec k^2_\perp,\uvec q_\perp^2)}\nn\\
&\quad\times\Bigg\{(1-x')^2\uvec k^2_\perp-(1-x)^2\uvec k'^2_\perp\nn\\
&\qquad+2(\uvec k_\perp\times\uvec k'_\perp)^2\left[\frac{(1-x)^2}{D(\uvec k^2_\perp)}-\frac{(1-x')^2}{D(\uvec k'^2_\perp)}-\frac{(1-x)(1-x')}{D(\uvec k^2_\perp,\uvec q_\perp^2)}\right]\Bigg\}
\end{align}
with $D(\uvec k'^2_\perp)=\uvec k'^2_\perp+\Lambda^2(x',1)$. Performing the integrals over the transverse momenta in $d=2-2\epsilon$, we find again that the $1/\epsilon^2$-divergences cancel out, and so at best
\begin{equation}
\left[-i\epsilon^{ij}_\perp\nabla^i_\perp\mathcal M^j_{C}(\Delta)\right]_{\Delta=0}=\mathcal O(1/\epsilon).    
\end{equation}


\begin{thebibliography}{99}

\bibitem{Accardi:2012qut}
  A.~Accardi {\it et al.},
  Eur.\ Phys.\ J.\ A {\bf 52} (2016) no.9,  268.

\bibitem{Proceedings:2020eah}
  A.~Prokudin, Y.~Hatta, Y.~Kovchegov and C.~Marquet,
``Proceedings, Probing Nucleons and Nuclei in High Energy Collisions: Dedicated to the Physics of the Electron Ion Collider: Seattle (WA), United States, October 1 - November 16, 2018,''
arXiv:2002.12333 [hep-ph].
  
\bibitem{Jaffe:1989jz}
  R.~Jaffe and A.~Manohar,
  Nucl.\ Phys.\ B \textbf{337} (1990) 509-546.

\bibitem{Ji:1996ek}
  X.~Ji,
  Phys.\ Rev.\ Lett.\  \textbf{78} (1997) 610-613.

\bibitem{Leader:2013jra}
  E.~Leader and C.~Lorc\'e,
  Phys.\ Rept.\  \textbf{541} (2014) no.3, 163-248.

\bibitem{Wakamatsu:2014zza}
  M.~Wakamatsu,
  Int.\ J.\ Mod.\ Phys.\ A \textbf{29} (2014) 1430012.

\bibitem{Wakamatsu:2010qj}
  M.~Wakamatsu,
  Phys.\ Rev.\ D {\bf 81} (2010) 114010.

\bibitem{Burkardt:2012sd}
  M.~Burkardt,
  Phys.\ Rev.\ D {\bf 88} (2013) no.1,  014014.

\bibitem{Burkardt:2003yg}
  M.~Burkardt,
  Phys.\ Rev.\ D {\bf 69} (2004) 057501.

\bibitem{Burkardt:2004ur}
  M.~Burkardt,
  Phys.\ Rev.\ D {\bf 69} (2004) 091501.
  
  \bibitem{Ji:2015sio}
  X.~Ji, A.~Schäfer, F.~Yuan, J.~H.~Zhang and Y.~Zhao,
  Phys.\ Rev.\ D {\bf 93} (2016) no.5,  054013.

\bibitem{Engelhardt:2017miy}
  M.~Engelhardt,
  Phys.\ Rev.\ D {\bf 95} (2017) no.9,  094505.

\bibitem{Engelhardt:2020qtg}
M.~Engelhardt, J.~R.~Green, N.~Hasan, S.~Krieg, S.~Meinel, J.~Negele, A.~Pochinsky and S.~Syritsyn,
Phys. Rev. D \textbf{102} (2020), 074505.

\bibitem{Hatta:2019csj}
  Y.~Hatta and X.~Yao,
  Phys.\ Lett.\ B {\bf 798} (2019) 134941.

\bibitem{Wakamatsu:2017isl}
  M.~Wakamatsu, Y.~Kitadono and P.-M.~Zhang,
  Annals Phys.\  {\bf 392} (2018) 287.

\bibitem{Burkardt:2002ks}
  M.~Burkardt,
  Phys.\ Rev.\ D {\bf 66} (2002) 114005.

\bibitem{Burkardt:2003uw}
  M.~Burkardt,
  Nucl.\ Phys.\ A {\bf 735} (2004) 185.

\bibitem{Burkardt:2003je}
  M.~Burkardt and D.~S.~Hwang,
  Phys.\ Rev.\ D {\bf 69} (2004) 074032.

\bibitem{Bacchetta:2011gx}
  A.~Bacchetta and M.~Radici,
  Phys.\ Rev.\ Lett.\  {\bf 107} (2011) 212001.

\bibitem{Gamberg:2009uk}
  L.~Gamberg and M.~Schlegel,
  Phys.\ Lett.\ B {\bf 685} (2010) 95.

\bibitem{Pasquini:2019evu}
  B.~Pasquini, S.~Rodini and A.~Bacchetta,
  Phys.\ Rev.\ D {\bf 100} (2019) no.5,  054039.
  
  \bibitem{Brodsky:2002cx}
  S.~J.~Brodsky, D.~S.~Hwang and I.~Schmidt,
  Phys.\ Lett.\ B {\bf 530} (2002) 99.

\bibitem{Chen:2008ag}
  X.~S.~Chen, X.~F.~Lu, W.~M.~Sun, F.~Wang and T.~Goldman,
  Phys.\ Rev.\ Lett.\  {\bf 100} (2008) 232002.

\bibitem{Wakamatsu:2010cb}
  M.~Wakamatsu,
  Phys.\ Rev.\ D {\bf 83} (2011) 014012.
  
\bibitem{Lorce:2012rr}
  C.~Lorc\'e,
  Phys.\ Rev.\ D {\bf 87} (2013) no.3,  034031.

\bibitem{Lorce:2013gxa}
  C.~Lorc\'e,
  Phys.\ Rev.\ D {\bf 88} (2013) 044037.

\bibitem{Lorce:2013bja}
  C.~Lorc\'e,
  Nucl.\ Phys.\ A {\bf 925} (2014) 1.
  
\bibitem{Boer:1997bw}
  D.~Boer, P.~J.~Mulders and O.~V.~Teryaev,
  Phys.\ Rev.\ D {\bf 57} (1998) 3057.
  
\bibitem{Boer:2003cm}
  D.~Boer, P.~J.~Mulders and F.~Pijlman,
  Nucl.\ Phys.\ B {\bf 667} (2003) 201.
  
\bibitem{Hatta:2011ku}
  Y.~Hatta,
  Phys.\ Lett.\ B {\bf 708} (2012) 186.

\bibitem{Lorce:2012ce}
  C.~Lorc\'e,
  Phys.\ Lett.\ B {\bf 719} (2013) 185.
  
\bibitem{Lorce:2015lna}
  C.~Lorc\'e,
  JHEP {\bf 1508} (2015) 045.
    
\bibitem{Ji:2002aa}
  X.~d.~Ji and F.~Yuan,
  Phys.\ Lett.\ B {\bf 543} (2002) 66.
  
\bibitem{Mulders:1995dh}
  P.~Mulders and R.~Tangerman,
  Nucl. Phys. B \textbf{461} (1996) 197-237.

\bibitem{Diehl:2015uka}
  M.~Diehl,
  Eur. Phys. J. A \textbf{52} (2016) no.6, 149.

\bibitem{Meissner:2007rx}
  S.~Meissner, A.~Metz and K.~Goeke,
  Phys.\ Rev.\ D {\bf 76} (2007) 034002.

\bibitem{Collins:2002kn}
  J.~C.~Collins,
  Phys.\ Lett.\ B {\bf 536} (2002) 43.
  
\bibitem{Sivers:1989cc}
  D.~W.~Sivers,
  Phys.\ Rev.\ D {\bf 41} (1990) 83.

\bibitem{Antille:1980th}
  J.~Antille, L.~Dick, L.~Madansky, D.~Perret-Gallix, M.~Werlen, A.~Gonidec, K.~Kuroda and P.~Kyberd,
  Phys.\ Lett.\ B {\bf 94} (1980) 523.

\bibitem{Airapetian:2004tw}
  A.~Airapetian {\it et al.} [HERMES Collaboration],
  Phys.\ Rev.\ Lett.\  {\bf 94} (2005) 012002.

\bibitem{Alexakhin:2005iw}
  V.~Y.~Alexakhin {\it et al.} [COMPASS Collaboration],
  Phys.\ Rev.\ Lett.\  {\bf 94} (2005) 202002.

\bibitem{Alekseev:2010rw}
  M.~G.~Alekseev {\it et al.} [COMPASS Collaboration],
  Phys.\ Lett.\ B {\bf 692} (2010) 240.

\bibitem{Aghasyan:2017jop}
  M.~Aghasyan {\it et al.} [COMPASS Collaboration],
  Phys.\ Rev.\ Lett.\  {\bf 119} (2017) no.11,  112002.

\bibitem{Boer:2015vso}
D.~Boer, C.~Lorc\'e, C.~Pisano and J.~Zhou,
Adv. High Energy Phys. \textbf{2015} (2015), 371396.

\bibitem{Davydychev:1993pg}
  A.~I.~Davydychev, V.~A.~Smirnov and J.~B.~Tausk,
  Nucl.\ Phys.\ B {\bf 410} (1993) 325.

\bibitem{Weinberg:1959nj}
  S.~Weinberg,
  Phys.\ Rev.\  {\bf 118} (1960) 838.

\bibitem{Smirnov:1990rz}
  V.~A.~Smirnov,
  Commun.\ Math.\ Phys.\  {\bf 134} (1990) 109.

\bibitem{Kang:2010hg}
  Z.~B.~Kang, J.~W.~Qiu and H.~Zhang,
  Phys.\ Rev.\ D {\bf 81} (2010) 114030.

\bibitem{Qiu:2020oqr}
  J.~W.~Qiu, T.~C.~Rogers and B.~Wang,
  Phys.\ Rev.\ D \textbf{101} (2020) no.11, 116017.

\bibitem{Goeke:2006ef}
  K.~Goeke, S.~Meissner, A.~Metz and M.~Schlegel,
  Phys.\ Lett.\ B {\bf 637} (2006) 241.

\bibitem{Bakker:2004ib}
  B.~L.~G.~Bakker, E.~Leader and T.~L.~Trueman,
  Phys.\ Rev.\ D {\bf 70} (2004) 114001.

\bibitem{Lorce:2011ni}
  C.~Lorc\'e, B.~Pasquini, X.~Xiong and F.~Yuan,
  Phys.\ Rev.\ D {\bf 85} (2012) 114006.
  
\bibitem{Burkardt:2008ua}
  M.~Burkardt and H.~BC,
  Phys.\ Rev.\ D {\bf 79} (2009) 071501.

\bibitem{Lorce:2017wkb}
  C.~Lorc\'e, L.~Mantovani and B.~Pasquini,
  Phys.\ Lett.\ B {\bf 776} (2018) 38.
  
\bibitem{Lorce:2018zpf}
  C.~Lorc\'e,
  Eur.\ Phys.\ J.\ C {\bf 78} (2018) no.9,  785.

\bibitem{Belitsky:2002sm}
  A.~V.~Belitsky, X.~Ji and F.~Yuan,
  Nucl.\ Phys.\ B {\bf 656} (2003) 165.

\bibitem{Lorce:2017isp}
  C.~Lorcé,
  Phys.\ Rev.\ D {\bf 97} (2018) no.1,  016005.

\end{thebibliography}
\end{document}